# Transforming Siliconization into Slippery Liquid-like Coatings

*Hernán Barrio-Zhang[1$], Glen McHale[1*], Gary G. Wells[1], Rodrigo Ledesma-Aguilar[1], Rui Han[2], Nicholas Jakubovics[3], Jinju Chen[2].*

[1]Institute for Multiscale Thermofluids, School of Engineering, University of Edinburgh, The King's Buildings, Mayfield Road, Edinburgh EH9 3FB, UK. [2]Department of Materials, Loughborough University, Loughborough, LE11 3TU, UK. [3]School of Dental Sciences, Faculty of Medical Sciences, Newcastle University, Newcastle Upon Tyne, NE2 4BW, UK.

*Email: glen.mchale@ed.ac.uk

$Email: hbarrio@exseed.ed.ac.uk

**ABSTRACT**

Siliconization is widely used as a coating technique to engineer surface properties, such as in the pharmaceutical and medical device industries to lubricate motion, ensure complete dispensation of product, and to inhibit protein adsorption and biofilm growth. In the hitherto unconnected literature, there has recently been significant progress in understanding the concept of surfaces slippery to liquids. Whereas in the siliconization industry the wettability of surfaces focuses on the hydrophobicity, as measured by contact angle and surface energy, for surfaces slippery to liquids the focus is on the contact angle hysteresis (droplet-on-solid static friction). Moreover, it has been discovered that surfaces with similar static wetting properties can have dramatically different droplet kinetic friction. Here, we report a simple-to-apply coating method to create ultra-low contact angle hysteresis liquid-like coatings for glass (G), polydimethylsiloxane (PDMS), polyurethane (PU) and stainless steel (SS); materials which are used for pharmaceutical/parenteral packaging and medical equipment. Moreover, we demonstrate that the coating's slow sliding dynamics surface properties for water droplets, which indicates high droplet kinetic friction, can be converted into fast sliding dynamics, which indicates low droplet kinetic friction, by a simple molecular capping (methylation) process. Our results provide new insight into key aspects of siliconization coatings in the context of industrial/commercial processes.

1. INTRODUCTION

Siliconization of surfaces is a term used in the pharmaceutical and medical device industry to refer to the coating of surfaces with a medical grade silicone oil or dimethicone (polydimethylsiloxane) (PDMS) layer[1] (also see various company literature[2–5]). This process is commonly used as a protective or barrier coating to minimize interactions of products and their container surfaces, such as glass containers for drug formulations, pre-filled syringes, syringe barrels and medical instruments implanted or inserted into the human body. Silicones are particularly important because of their biocompatibility. In the medical device and pharmaceutical siliconization literature, the focus in surface wettability is on the hydrophobicity of the siliconized surfaces and on the lubricating properties to ensure complete dispensation/drainage of product without leaving residue in containers and to lubricate motion between solid surfaces or for insertion of medical devices reducing gliding forces, e.g., in syringe barrels[6–8] and catheters. Siliconization is also considered an important aspect of preventing biofilm formation due to its anti-protein adsorption properties. This is particularly important in healthcare, where biofilms are a major source of post-operative infections, resulting in thousands of deaths in healthcare settings [9–11] However, the emphasis in industrial/commercial siliconization processes on hydrophobicity does not address concepts of how easily a contact line may be set into motion or how fast a liquid may be shed from the surface, which are properties dependent on the contact line pinning and dynamic contact line friction.[12–15]

Two important types of siliconization coatings are oily-siliconization, and baked-on siliconization for industrial production lines, and organopolysiloxane siliconizing fluids, typically for use with laboratory glassware. Oily siliconization of glassware is often performed in industry by spraying methyl-terminated PDMS with viscosities of 1000-12500 cSt or commercial medical grade silicone oil emulsions, such as Dow Corning® 365/366, 35% Dimethicone NF Emulsion (DC 365/366), using automated units. A subsequent dry-heat oven treatment at ca 220-300 °C for, e.g., 10-30 minutes is often used to produce baked-on siliconization. Alternatively, siliconization can be performed by wiping or dip coating followed by drying at room or elevated temperature. Despite the extensive use of silicone oil coatings in pharmaceutical containers and pre-filled syringes, there are concerns about the interaction of protein molecules within product formulations with the oil coated surfaces and of the migration of silicone oil into the product.[16–18]

A third type of commercially available siliconization uses siliconizing fluids based on organopolysiloxanes, such as Sigmacote® which consists of 1-3% 1,7-Dichloro-1,1,3,3,5,5,7,7-octamethyltetrasiloxane in heptane, or products such as Aquasil™ or Surfasil™. These are immersion or wipe coated onto glass, ceramics, plastics and some metal surfaces, and interact with the native hydroxyls (-OH) on the surfaces to form a covalently-attached hydrophobic PDMS coating that retards clotting of blood or plasma and resists adsorption of many basic protein bindings[19]. None of the three approaches to siliconization described above emphasize hydroxylation (activation) of surfaces prior to coating. Moreover, all tend to focus on the hydrophobic properties of the coated surfaces *via* surface energy[1,20] and contact angle measurements[1,21,22] as the key surface wetting property. They often do not report the contact angle hysteresis and never report dynamic contact line friction properties.

Independent of the development of industrial/commercial siliconization techniques within the health/biomedical context, there has been significant progress over the last decades on the understanding of surfaces slippery to liquids. One of the first approaches was to combine surface structure or porosity with a hydrophobic material or coating to create a superhydrophobic surface ("Lotus effect").[23–26] In this case, droplets of water ball-up with contact angles approaching 180º and roll off the surface. However, motivated by a desire to avoid pressure-induced collapse of the superhydrophobic state into the surface structure, which results in pinned droplets, slippery liquid-infused porous surfaces (SLIPS)[27] and slippery lubricant-impregnated surfaces (LIS)[28] were invented. In parallel with these developments, methods of covalently attaching short PDMS (or other) polymer chains with glass transition temperatures below room temperature to surfaces to provide coatings with ultra-low contact angle hysteresis liquid-like properties, which avoid the risk of liquid lubricant depletion, were developed.[29–32]. Although silicone oil and PDMS-based liquid-infused/lubricant impregnated surfaces and liquid-like surfaces are hydrophobic, the emphasis for liquid-like coatings is on coatings giving ultra-low contact angle hysteresis (the difference between the advancing and receding contact angles, $\Delta\theta_{CAH} = \theta_A - \theta_R$). This is because the in-plane liquid-on-solid static friction a droplet (or a contact line) needs to overcome to induce motion is proportional to $\Delta\theta_{CAH}F_N$ where $F_N$ is the normal component of the surface tension force and depends on the equilibrium contact angle, $\theta_e$.[13] Both hydrophobic and hydrophilic surfaces can be slippery to liquids if the contact angle hysteresis is ultra-low, because this decouples the normal surface tension force from the in-plane droplet-on-substrate friction.

The cross-over between industrial/commercial siliconization techniques and progress on understanding and making surfaces slippery to liquids is very limited in the literature. For example, oily-siliconization could be viewed as similar to creating SLIPS surfaces for which we have recently shown how to define a liquid version of Young's law valid for predicting contact angles of droplets on thin lubricant layers.[33] Similarly, baked-on siliconization has a direct analogue to the creation of covalently-attached liquid-like coatings by thermal grafting of silicone oil coatings on substrates[29,34], for which it is known hydroxylation (activation) of the surface and choice of silicone oil viscosity are important for homogeneous surfaces with ultra-low contact angle hysteresis.[34] Finally, the use of siliconizing fluids is similar to liquid-like surface coatings, such as quasi-liquid surfaces (QLS) [35–38] which use a chlorosilane approach and slippery omniphobic covalently-attached liquid surfaces (SOCAL[30]) which use a methoxysilane approach, for which we have shown hydroxylation (activation) of surfaces and optimization for ultra-low contact angle hysteresis are critical.[39] In addition, the literature on slippery surfaces has identified that surfaces may have similar static wetting properties and droplet-on-solid static friction, but very different droplet-on-solid kinetic friction.[14] The origin of this difference in kinetic friction is the interaction between water and hydroxy-terminated PDMS chains, typical of chlorosilane and methoxysilane-based coating methods, compared to interactions with methyl-terminated PDMS coatings, typical of thermally-grafted silicone oil coatings. It is now possible to convert surfaces from high to low droplet-on-solid kinetic friction by a molecular capping (methylation) procedure of the hydroxy-terminated PDMS coatings.[15]

Here, we report the use of a siliconization fluid Sigmacote® to create liquid-like coatings with ultra-low contact angle hysteresis on glass (G), polydimethylsiloxane (PDMS), polyurethane (PU) and stainless steel (SS), as exemplars of pharmaceutical and medically-relevant surfaces. To do so, we incorporate an activation of surfaces inspired by liquid-like surface coatings, use a spray-on technique inspired by industrial baked-on siliconization, and optimize the coating process for ultra-low contact angle hysteresis rather than hydrophobicity. We also consider the use of Sigmacote® as a base coating onto which a subsequent liquid-like coatings (SOCAL and QLS) can be added. Finally, we show the droplet-on-solid kinetic friction on ultra-low contact angle hysteresis Sigmacote®-coated substrate, and on liquid-like coatings applied to a Sigmacote®-coated substrate, can be converted to a significantly lower droplet-on-solid kinetic friction by a final molecular capping (methylation) step resulting in a significantly faster sliding of water droplets (or equivalently of three-phase contact lines).

## 2. DROPLET-ON-SOLID FRICTION AND LIQUID-LIKE SURFACE CONCEPTS

Young's law[40] states that the balance of interfacial forces at the three-phase contact line of a liquid-solid-gas system will form a unique static configuration defined by an equilibrium contact angle, $\theta_e$,

$$\cos \theta_e = \frac{\gamma_{SG} - \gamma_{SL}}{\gamma_{LG}} \tag{1}$$

where $\gamma_{SG}$ is the interfacial tension of the solid-gas interface, $\gamma_{SL}$ is the interfacial tension of the solid-liquid interface and $\gamma_{LG}$ is the surface tension of the liquid-gas interface. However, Young's law applies to the ideal situation where a surface is perfectly smooth and chemically homogeneous. In contrast, practical surfaces tend not to be topographically smooth or chemically uniform[26]. Thus, pinning forces exist which allow a droplet to adopt a window of sessile states and result in an observed static contact angle, $\theta_s$, different to the equilibrium contact angle, $\theta_e$. The range of possible values of static contact angles provides information on the lateral in-plane forces required to initiate motion of a three-phase contact line or a droplet on a surface. This range is characterized by two delimiting values, the advancing and receding contact angles, $\theta_A$ and $\theta_R$, respectively, with the interval of possible static contact angles defined as the contact angle hysteresis ($\Delta\theta_{CAH}$),

$$\Delta\theta_{CAH} = \theta_A - \theta_R. \tag{2}$$

The concept of contact angle hysteresis is important because the lateral frictional pinning force for a droplet on the surface, $F_p$, is directly proportional to the value of $\Delta\theta_{CAH}$ and scales with the normal component of the surface tension force, $F_N$,[13] i.e.,

$$F_p = \left(\frac{k\Delta\theta_{CAH}}{\pi}\right) F_N \tag{3}$$

where, for a droplet with a circular contact area of radius $r_c$, $k=\pi/4$ and $F_N = 2\pi r_c \gamma_{LG} \sin\theta_e$. From a siliconization perspective focused on shedding water from the coated surface, this means variations in the contact angle, e.g. from ~90° to 110°, are far less important than variations in the contact angle hysteresis – if the contact angle hysteresis doubles, the pinning force also doubles. The contact angle hysteresis represents the effect of chemical heterogeneity and/or roughness of the surface and translates the liquid adhesion due to the equilibrium contact angle into static liquid-on-solid friction causing the three-phase contact line and droplets to pin. The key difference between the PDMS-based liquid-like surface literature and industrial/commercial siliconization is the change in focus from optimizing hydrophobicity through the contact angle to optimizing low contact line pinning and droplet-on-solid static friction by minimization of contact angle hysteresis.

In the liquid-like surfaces literature it is generally recognized that a surface is ultra-smooth, and hence slippery to liquids, if the contact angle hysteresis is below 5°.[32] Interestingly, ultra-smooth liquid-like surfaces that exhibit low in-plane friction to liquids have been recognized to inhibit the formation of biofilms. For example, recent work conducted by Zhu *et al*[41] has shown that ultra-smooth surfaces are promising as an antibiofilm strategy in both static and flow conditions. These surfaces are unique for they present unprecedented levels of liquid repellence through the minimal levels of contact line pinning, which is attributed to the flexible nature of the polymer chains grafted onto them.[30] Liquid-like surfaces are produced by selecting a short polymer chain with a glass-transition temperature below room temperature (to ensure chain flexibility) (Figure 1a), activating (hydroxylating) the target substrate surface (Figure 1b), and covalently attaching the chains to the substrate without cross-linking (to ensure chain flexibility is not lost) (Figure 1c)[31]; the average coating thicknesses are typically ~5 nm[42] although a thickness of as much as 30 nm has been reported.[35] The specific wettability and contact line pinning (static liquid friction) properties of the surface depend on the length of the chains and the type of polymer used to coat the surface. The low contact line static friction properties of liquid-like surfaces are believed to originate from the flexibility of the attached polymer chains in mushroom-like rather than brush-like conformations.

Some notable liquid-like surface coatings are Slippery Omniphobic Covalently-Attached Liquid (SOCAL) surfaces ($\Delta\theta_{CAH} \sim 1-3°$) [30], Quasi-Liquid Surfaces (QLS) ($\Delta\theta_{CAH} \sim 3°$) [36,37,43], thermally-grafted silicone-oil (TGSO) coatings[29,34], for, and PEGylated surfaces ($\Delta\theta_{CAH} \sim 2-3°$)[44–46]. The first three of these types of surfaces are all examples using PDMS chains and can be categorized as siliconization techniques. However, SOCAL and QLS result in hydroxy-terminated PDMS coatings whilst TGSO results in a methyl-terminated PDMS coating. All three types of surfaces have similar hydrophobic values of static contact angles for water, typically 100°-108°. Furthermore, advances have shown that hydroxy-terminated PDMS liquid-like surfaces can display low values of contact angle hysteresis (i.e. droplet-on-solid static friction) while having low mobility of droplets when they are in motion, i.e. they can retain relatively large values of droplet kinetic friction.[14,15] Recently, studies have shown that it is possible to reduce the droplet kinetic friction on hydroxyl-terminated PDMS-based liquid-like surfaces by converting the PDMS chains to methyl-terminations using a molecular capping (methylation) procedure.[14,15] In the remainder of this paper, we focus on a novel deposition technique using a siliconization fluid Sigmacote®, which results in ultra-low droplet friction liquid-like coatings with high or low droplet kinetic friction on four medically and industrially-relevant surfaces.

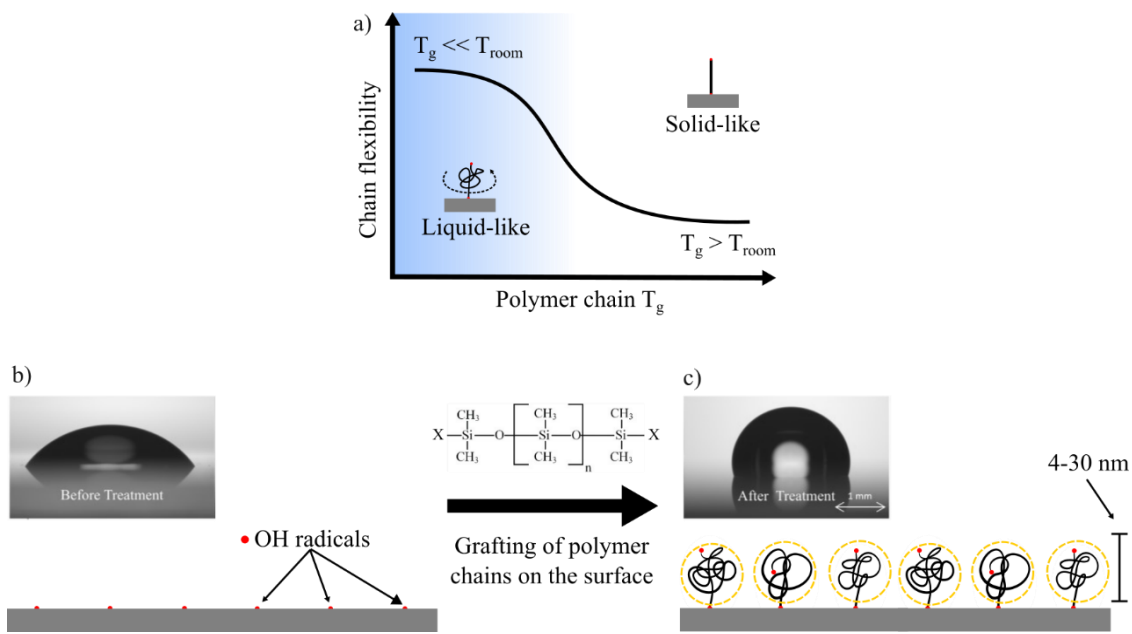

**Figure 1.** PDMS-based liquid-like coatings. (a) Glass transition temperature for grafted polymer chains (adapted with permission from reference[14]). Under room temperature conditions, grafted polymer chains adopt flexible, liquid-like behaviour. The red dots indicate sites of OH terminals. (b) Shows a hydroxylated (activated) surface which has hydroxyl (OH) sites for covalent attachment; the activated surface is hydrophilic. To produce liquid-like surfaces, the activated surface is reacted with methylpolysiloxanes with different lengths and terminations (marked by X) *via* different salination strategies. (c) The reaction grafts mushroom-like polymer chains of different lengths onto the surface, achieving low static droplet friction and low wettability for water

## 3. EXPERIMENTAL SECTION

**Substrates.** The substrates used were standard 25×75 mm soda-lime microscope glass slides (G), 10:1 polymer and crosslinker ratio for PDMS, standard 6-mm polyurethane (PU) tubing and 0.5-mm thick 304S15 stainless-steel (SS) cut into 20×70 mm samples.

**Standard Coating Procedure (SCP).** Sigmacote® is a chlorinated organopolysiloxane in heptane that produces a neutral, hydrophobic film, which repels water, retards the clotting of blood or plasma, and prevents surface adsorption of many basic proteins. The reagent forms a covalent bond on glass surfaces (and other surfaces such as ceramics) - the chlorosilane ingredient reacts with the -OH group on the glass surface to form this linkage. The coating is very robust and does not degrade easily. The surface to be siliconized must be clean and dry, here we used acetone and isopropanol (IPA). The surface is then covered or immersed in undiluted Sigmacote® and the excess is removed, resulting in an almost instantaneous reaction with the native hydroxyl groups present on the surface. The coated surface is then left to air dry in a fumehood at room temperature. The siliconized surface is then rinsed with water to remove any by-product HCl before use. An additional step to achieve a more durable coating consists in heating the surface at 100°C for 30 min after the heptane has evaporated.

**Cleaning Substrates.** All substrates, except those used in the standard coating procedure, were cleaned using the following procedure. First, we immerse the substrate in a 2% v/v solution of Decon90 detergent and de-ionized (DI) water, then sonicate it for 10 minutes in an ultrasonic bath. Next, we rinse the substrate with DI water and sonicate it again in pure DI water for 10 minutes to remove any remaining cleaning agent. After rinsing, we place the substrate in an acetone bath for a 10-minute sonication, repeating this step in an IPA bath. Finally, we air-dry the substrate, place it in a petri dish to await plasma activation.

**Surface Activation**. Depositing hydroxyl groups on the substrate (surface activation) is essential for creating liquid-like coatings. To activate a clean surface, we place the sample inside a plasma oven (Henniker HPT-200) and expose it to oxygen plasma at 15 sscm, maintaining an environmental pressure below 0.9 mbar. To obtain the optimized coating parameters for ultra-low contact angle hysteresis on each surface, plasma power and plasma exposition sweeps were conducted.[39] For plasma powers, the range explored was between 60 W, 120 W and 200 W (Max power) while plasma activation times where varied between 5 to 30 minutes in increments of 5 mins. The parameters were considered optimal if the resulting contact angle hysteresis of the surface was consistently less than 2°. The optimized parameters resulting from this study are given in Table 1.

Table 1. Substrate specific parameters for surface activation prior to coating.

| SUBSTRATE | POWER (W) | PLASMA TIME (MIN) |
|---|---|---|
| GLASS | 200 | 10 |
| POLYDIMETHYLSILOXANE | 200 | 5 |
| POLYURETHANE | 120 | 20 |
| STAINLESS STEEL | 200 | 10 |

**Spray-on Sigmacote® Liquid-like Coating (SoSLIC).** Figure 2 illustrates the general procedure for depositing a liquid-like surface using Sigmacote®. Immediately after surface activation, we place the sample flat in the fume hood and use an air-spray gun to apply the first coating layer. The sample dries for 30 minutes before we spray on a second layer, and then repeat this step a third time. We then let the sample rest for an additional hour before rinsing with DI water, IPA, and toluene. For stainless steel, we add a final step to enhance surface durability by placing the coated sample on a hot plate set to 100 °C for one hour.

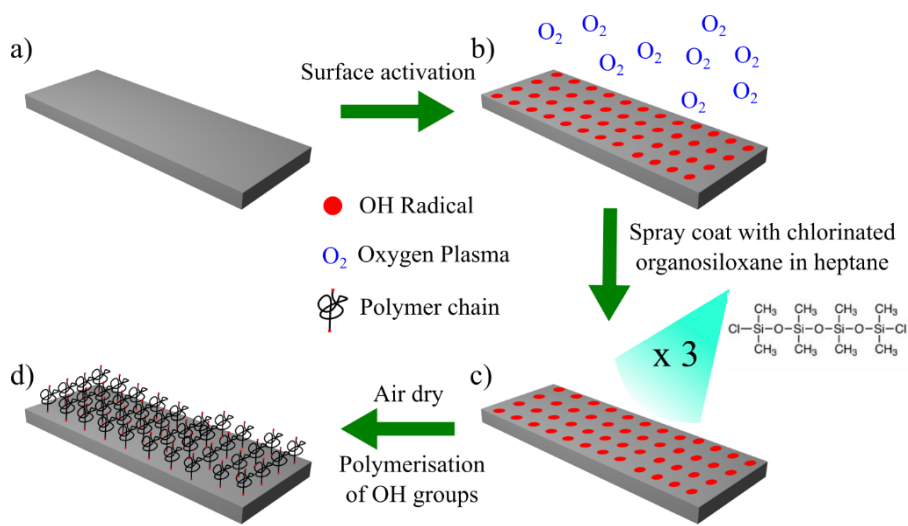

**Figure 2.** Spray-on liquid-like coating procedure. (a) A substrate is cleaned prior to activation. (b) The substrate is activated using an Oxygen ($O_2$) plasma. (c) Three layers of Sigmacote® are spray-coated onto the substrate. (d) The coated substrate is left for 1 hour to complete polymerizing.

**SOCAL Coating.** Immediately after surface activation, we hand dip the activated substrate into a solution of IPA, dimethyldimethoxysilane (Sigma-Aldrich), and sulfuric acid (100%, 10%, and 1% wt, respectively) for ~5 seconds, slowly withdrawing it and gently tapping it on absorbent material to remove any excess solution. Next, we place the sample in a bespoke humidity chamber set at 60% ± 1% relative humidity (RH) for 20 minutes, allowing the polycondensation reaction to occur. We then remove the sample, thoroughly rinse it with DI water, IPA, and toluene to eliminate any unreacted material, air-dry it, and place it inside a petri dish.

**QLS Coating.** Immediately after surface activation, we place the sample inside a glass Petri dish (radius = 10 mm) along with a small hourglass. Using a micropipette, we add 150 μL of 1,3-dichlorotetramethydisiloxane (Sigma-Aldrich, molecular weight 203.209 g/mol) onto the hourglass and quickly cover the dish with its lid. We allow the polycondensation reaction to proceed in the vapor phase for 5 minutes, then remove the sample and rinse it with DI water, IPA, and toluene. Finally, we store the sample in a Petri dish until further use.

**Molecular Capping (Methylation) of Surfaces.** The SOCAL and QLS coatings on the substrates are covalently-attached hydroxy-terminated PDMS chains, i.e. substrate-O-$[Si(CH_3)_2$-$O]_n$-H. To molecularly cap the coatings, we place the coated substrates inside a 10-mm Petri dish in the presence of 100 μL of methylchlorosilane at 40% RH. This converts the hydroxy-terminated PDMS chains to methyl-terminated PDMS chains, i.e. substrate-O-$[Si(CH_3)_2$-$O]_n$-$Si(CH_3)_3$. We allow the reaction take place for 2 hours before rinsing the samples with water, IPA and toluene. The same molecular capping procedure is used with the SoSLlC on glass.

**Characterisation of Surfaces.** The surfaces are characterized using standard droplet goniometry. First, we place an 8 μL DI water (Ultrapur, Supelco) droplet on the sample then increase the droplet volume by 4 μL at a flowrate of 8 μL/min. After resting for 20 seconds, we decrease the droplet volume by 4 μL at the same flowrate. We record the advancing and receding contact angles, $\theta_A$ and $\theta_R$, at the onset of motion of the base radius to determine the contact angle hysteresis, $\Delta\theta_{CAH}$. To compare wettability of the surfaces, we refer to the advancing contact angle because it allows us to compare wettability from a well-defined value, independent of contact angle hysteresis. Static contact angles quoted within the literature can be any value between the advancing and receding contact angle, and tend to the receding contact angle when a droplet is evaporating. We measure all parameters using an open-source droplet shape analyzer (pyDSA[14]), determining the droplet baseline and fitting a third-degree polynomial to its interface. Each measurement represents an average of at least three trials, with errors reported as the standard deviation between measurements.

**Characterisation of Sliding Dynamics.** We deposit a 20 μL droplet on a surface inclined between 5° − 20°, varied in increments of 5°. Once the droplet reaches the desired volume, we start recording at 30 fps and detach the droplet from the needle. We observe the droplet as it travels 1 cm down the surface, measuring its velocity only at terminal velocity, in the absence of acceleration, through a linear fit to the displacement of the centre of the droplet. We repeat each experiment at least three times.

**Characterisation of Film Thickness**. To determine the thickness of the liquid-like coatings, we made use of ellipsometry measurements using Film Sense (FS-1EX). The measurements were done using silicon wafers (procured with Inseto) coated with SOCAL and SoSLIC. These measurements were repeated 5 times for each sample.

4. RESULTS AND DISCUSSION

The two metrics we use to determine whether a hydrophobic liquid-like coating has been achieved are water contact angle for wettability and contact angle hysteresis for droplet pinning. Since Sigmacote® uses 1,7-Dichloro-octamethyltetrasiloxane, we expect the covalently-attached coating to be a hydroxy-terminated PDMS chain with contact angle values

~100° − 108° and contact angle hysteresis values below 3° and, hence, comparable to both SOCAL and QLS. From the literature, a surface with contact angle hysteresis below 5° is regarded as ultra-smooth. [47]

Figure 3 shows the advancing contact angle and contact angle hysteresis values of the glass, PDMS, polyurethane and stainless-steel substrates after cleaning, but prior to any coating. All the substrates in Figure 3a have different levels of wettability, where glass and polyurethane are hydrophilic ($\theta_A < 90°$) and PDMS and stainless steel are hydrophobic ($\theta_A > 90°$) when using the advancing contact angle (rather than static contact angle) above or below 90° as the definition of hydrophobic or hydrophilic. Furthermore, all the substrates have large values of contact angle hysteresis with the lowest contact angle hysteresis in the range $\Delta\theta_{CAH} \sim 10°$-$22°$ (Figure 3b). These values are far removed from ones for surfaces with ultra-low contact angle hysteresis characteristic of slippery liquid-like surfaces. After coating the substrates with Sigmacote® using the standard procedure, there is a more uniform set of contact angles observed, distributed around $\theta_A \sim 90°$ (Figure 3c). The advancing contact angles for glass and PDMS change sufficiently to suggest the coating has been effective in providing a siliconization. However, the coated polyurethane surface remains slightly hydrophilic and shows no significant change in the advancing contact angle. Similarly, there is only a small change in the advancing contact angle for stainless-steel. Contact angle hysteresis reduces for all substrate materials, but remains high in the range $\Delta\theta_{CAH} \sim 10°$-$22°$ and is only a slight reduction for glass (Figure 3d). Thus, although glass and PDMS have been siliconized, the coating is insufficiently homogeneous to achieve ultra-low contact angle hysteresis. From equation 3, the expected reduction in the pinning force for small droplets of water on glass, polyurethane and stainless steel is ~10%, 30% and 40%, respectively. In contrast, there is an expected increase in pinning force of ~15% for small droplets on PDMS arising from the reduction in average contact angle towards 90º with only a marginal reduction in contact angle hysteresis.

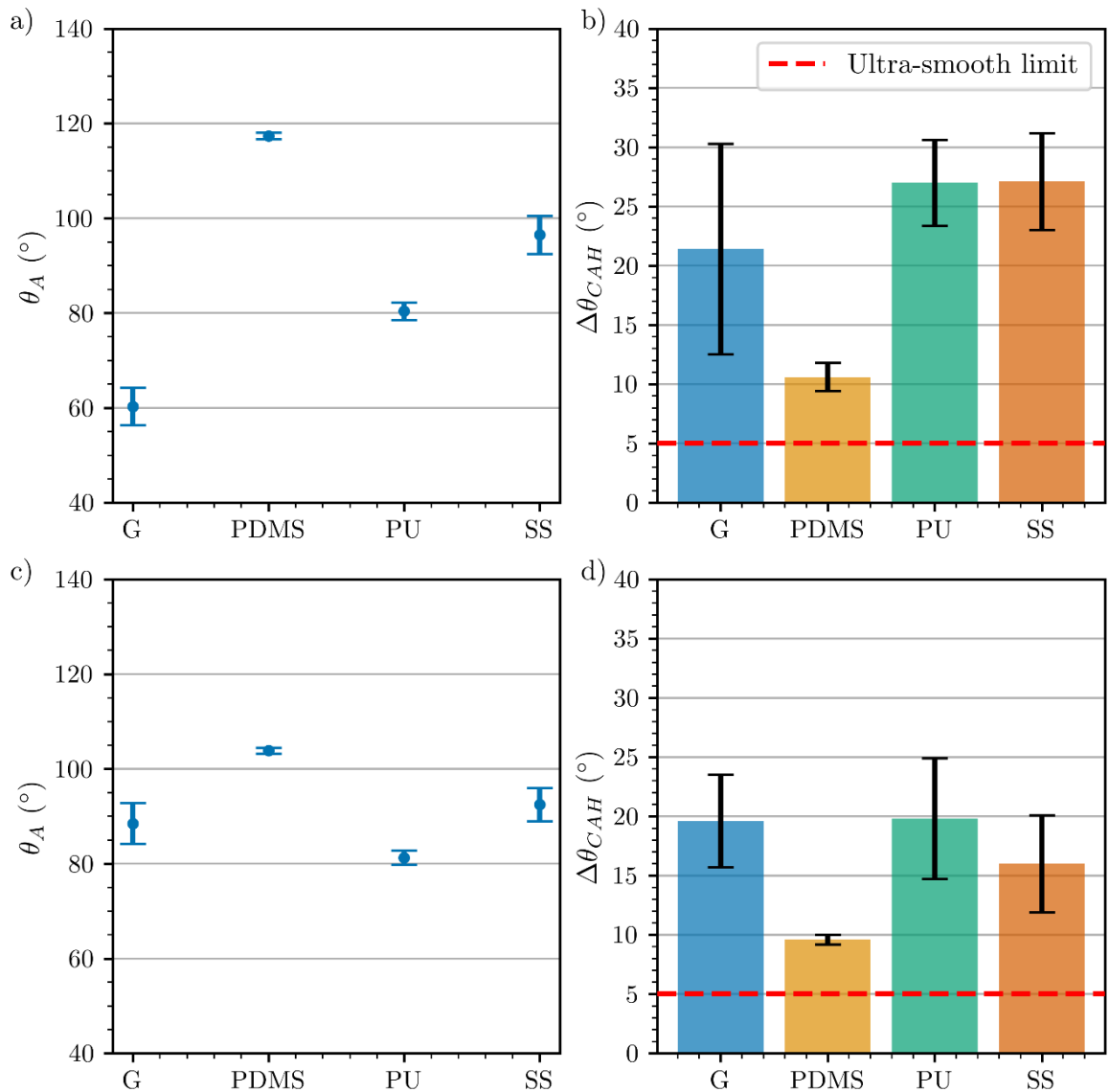

**Figure 3.** Wetting and contact angle hysteresis on cleaned, but uncoated, glass (G), PDMS, polyurethane (PU) and stainless-steel (SS) substrates and after the standard Sigmacote® coating. (a) Advancing contact angle (uncoated substrates). (b) Contact angle hysteresis (uncoated substrates). (c) Advancing contact angle (coated substrates). (d) Contact angle hysteresis (coated substrates). The red dotted line shows ultra-low contact angle hysteresis below 5º corresponding to the liquid-like surface criterion.[32]

To obtain slippery liquid-like surfaces, we developed a spray coat procedure to apply the Sigmacote® siliconization fluid (Figure 2) and investigated different oxygen plasma power exposure times until consistent results were achieved for ultra-low contact angle hysteresis $\Delta\theta_{CAH} < 3°$ (Table 1). The advancing contact angles achieved on the four substrate materials show a wide distribution around 100° (Figure 4a). For glass, the advancing contact angle of $\theta_A \sim 103°$ and contact angle hysteresis $\Delta\theta_{CAH} \sim 2°$ are both consistent with expectations from SOCAL and QLS coatings with covalently-attached hydroxy-terminated PDMS chains. For polyurethane, similar to the standard coating method, the advancing contact angle shows no significant change. In contrast, both PDMS and stainless-steel shows significant changes in advancing contact angle, but the former reduces to ~87° whilst the latter increases to ~123°. This suggests roughness enhancement of the contact angle for water droplets on stainless-steel. However, despite the lack of an advancing contact angle characteristic of a smooth layer of covalently-attached hydroxy-terminated PDMS chains, the contact angle hysteresis for all surfaces shows a dramatic decrease in contact angle to within the liquid-like threshold of ~3° (Figure 4b). From eq. 3 this is a significant reduction in the pinning force on small water droplets of ~90%-95% across all four surfaces. This reduction is almost entirely due to the reduction in the contact angle hysteresis with the pinning force being relatively insensitive to the changes in the average contact angle.

Ellipsometry measurements for the standard Sigmacote® coating method applied to silicon wafers gave an estimated layer thickness of 3.5±2.4 nm, which is significantly thinner than suggested in the datasheet. Although the coating is within the nanometric thickness typical for liquid-like coatings, the associated error suggests the coated layer is non-uniform and therefore consistent with the observation of relatively high contact angle hysteresis on the four substrate materials tested (glass, PDMS, polyurethane and stainless-steel) (Figure 2d). The ellipsometry measurements for our SoSLIC coating gave an estimate of 18.5±0.4 nm, which suggests good uniformity and is consistent with the ultra-low contact angle hysteresis reported for our SoSLIC-coated substrate materials (Figure 4b). This thickness is around three to four times that of typical liquid-like coatings, such as SOCAL which in our case we measured to be 6.3±0.3 nm.

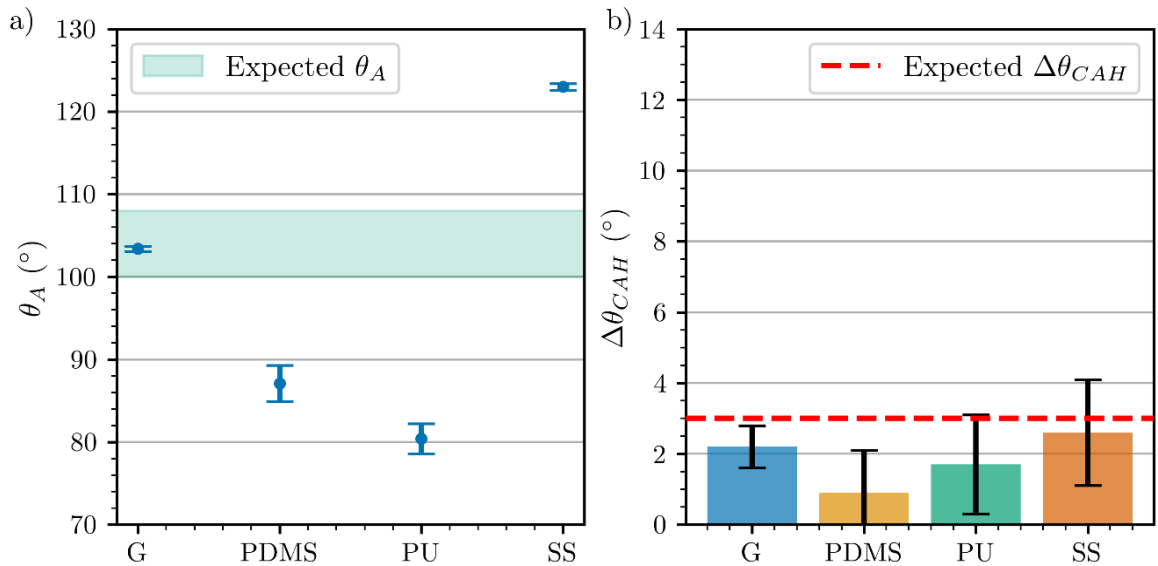

**Figure 4.** Wetting and contact angle hysteresis on glass (G), PDMS, polyurethane (PU) and stainless-steel (SS) substrates after oxygen plasma activation followed by spray-on Sigmacote® to create a SoSLIC surface. (a) Advancing contact angle; The shaded region indicates the advancing contact angle of PDMS-based liquid-like surfaces on glass. (b) Contact angle hysteresis; the red dotted line indicated the threshold contact angle hysteresis to satisfy our liquid-like surface criterion.

In the literature, attempts to create liquid-like surface coatings on a wide class of materials other than glass and silicon have included the use of a sol-gel coating as a first coating onto which the liquid-like coating can be subsequently covalently-attached.[48]. An alternative has been to use a sequence of three spray coating steps with an epoxy resin prepolymer with a curing agent, followed by polyamines to increase the number of reactive functional groups and, finally, a PDMS isopropanol solution.[49] We have also previously used a spin-on Sigmacote® coating followed by a SOCAL coating to convert an SU-8 photoresist to a slippery hydrophobic surface.[50] Here we investigated the possibility of using a spray-on method for a siliconizing solution Sigmacote® (i.e. SoSLIC), a room temperature coating, as a first coating layer on our four substrate materials (glass, PDMS, polyurethane and stainless-steel) onto which SOCAL and QLS liquid-like coatings could be subsequently be covalently-attached.

Figures 5a and 5b show advancing contact angles and contact angle hysteresis of the surfaces after the SOCAL deposition on all four substrate materials are both typical of SOCAL coated directly onto glass substrates. All four surfaces are hydrophobic with an ultra-low contact angle hysteresis below 3°. In the case of the QLS coating, we were able to succeed with glass, PDMS and polyurethane, but not stainless steel (Figures 5c and d) where the exposure to 1,3-dichlorotetramethydisiloxane produced visible damage to the coating (See Supplementary Information). The difference between SOCAL and QLS is that SOCAL uses an acid-catalyzed methoxysilane approach with methanol as the by-product, whereas QLS uses a self-catalyzed chlorosilane approach with hydrochloric acid as the by-product. In particular, the vapor phase polycondensation approach used for this particular approach to creating a QLS coating is likely to produce sufficient HCl to damage the Sigmacote® layer and/or the stainless-steel substrate. Interestingly, it also suggests the Sigmacote® can withstand the dilute sulphuric acid used as a catalyst in the SOCAL coating method.

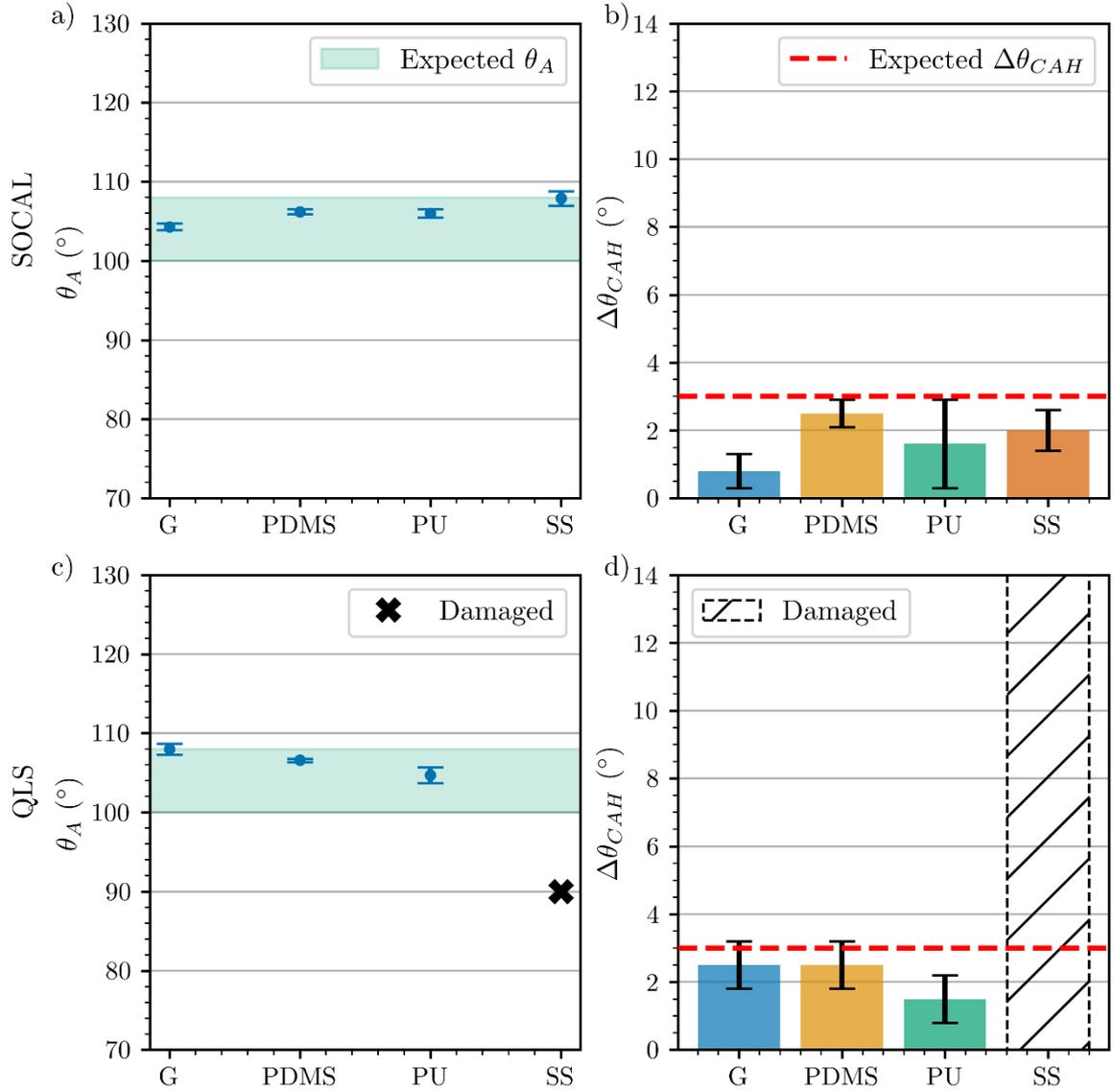

**Figure 5.** Wetting and contact angle hysteresis on glass, PDMS, polyurethane and stainless-steel substrates with a spray-on Sigmacote® coating followed by either a SOCAL or QLS liquid-like coating. (a) and (c) Advancing contact angle (SOCAL and QLS, respectively); The shaded region indicates the advancing contact angle of PDMS-based liquid-like surfaces on glass. (b) and (d) Contact angle hysteresis (SOCAL and QLS, respectively); the red dotted line indicated the threshold contact angle hysteresis to satisfy our liquid-like surface criterion. In panels (c) and (d), the coating on stainless steel was visibly damaged with high pinning behaviour after the coating process.

Finally, we considered whether the SoSLIC method results in a surface with dynamic behaviour similar to that reported for SOCAL[14], which has high droplet kinetic friction, even though the pinning force on the droplet (i.e. droplet static friction) is ultra-low.[14] To do so, we used a sliding droplet test whereby a droplet of a mass, $m$, achieving a steady speed, $v$, on a plane inclined at an angle $\alpha$ to the horizontal, will have a frictional force, $F_f$, opposing the component of the gravitational force down the incline driving its motion, i.e. $F_g = mg\sin\alpha$ where $g = 9.81$ ms$^{-2}$ is the acceleration due to gravity. A difference in kinetic friction for two drops of the same liquid and the same mass on two surfaces inclined at the same angle is visible from the difference in steady speeds the droplets adopt to achieve the same frictional force on the two surfaces.

In our experiments, we used 20 µL water droplets sliding at steady speeds on surfaces with tilt angles in the range 5° − 20° corresponding to frictional forces up to 70 µN. As the SoSLIC method creates a covalently-attached hydroxy-terminated PDMS coating, converting this to a methyl-terminated PDMS coating through molecular capping (methylation) should remove or reduce the interactions between the silanol groups and water which are believed to cause the high dynamic droplet friction.[14,15] Figure 6 shows a comparison between SOCAL and molecularly-capped SOCAL and the SoSLIC and molecularly-capped SoSLIC. For SOCAL, the molecular capping produces a ~55-fold increase in sliding speed of compared to the uncapped SOCAL surface (See supplementary). For the SoSLIC, the molecular capping produces a ~4-fold increase in sliding speed. However, whilst the relative change on molecularly capping SoSLIC is less dramatic than occurs when molecularly capping SOCAL, the speed of droplets on molecularly capped SoSLIC is ~ 4 times faster than molecularly capped SOCAL. It is unclear why this is the case, but one possibility might be the three-fold greater thickness of the SoSLIC coating (as measured by ellipsometry).

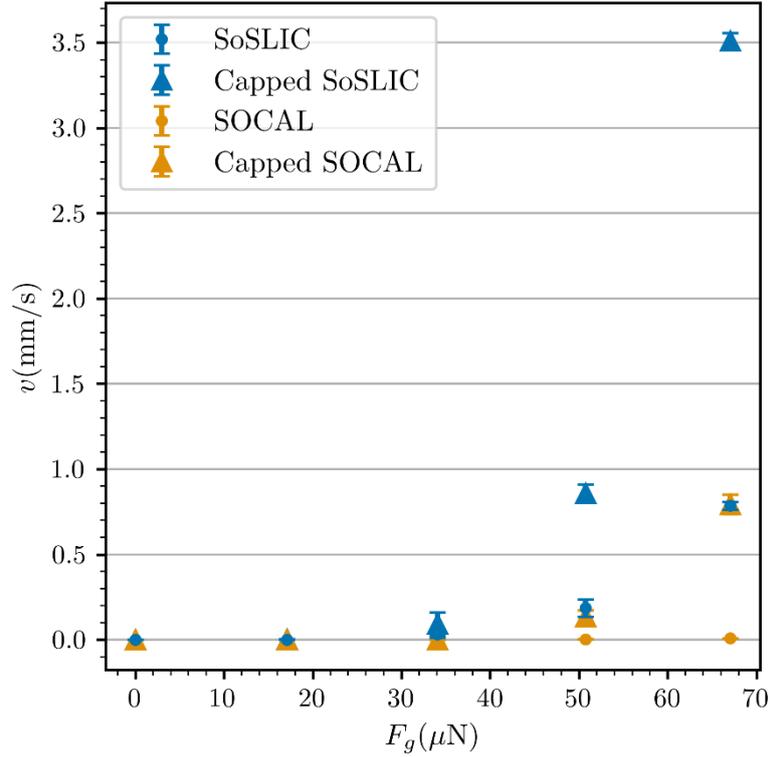

**Figure 6.** Steady state sliding speeds, *v*, of 20 μl water droplets on tilted SOCAL and spray-on Sigmacote ® coated (SoSLIC) glass substrates before and after molecularly capping (methylation). $F_g$ is the force due to gravity driving motion down surfaces inclined at angles up to 20º to the horizontal.

## 5. CONCLUSIONS

In this work, we started by discussing similarities between slippery liquid- and solid-lubricated ultra-low contact angle hysteresis surfaces and three siliconization techniques used in industry. From this we developed a spray-on siliconization process using surface activation (hydroxylation) and optimization of the coating process focusing on minimizing contact angle hysteresis, rather than hydrophobicity, to achieve liquid-like coatings. We demonstrated this could be applied to activated surfaces of glass, polydimethylsiloxane, polyurethane and stainless steel (materials relevant to the pharmaceutical/parenteral packaging and medical equipment) to create slippery ultra-low contact angle hysteresis (< 3º) coatings. We have also shown that the mobility of contact lines and water droplets, which can be significantly improved through a molecular capping (methylation) step, is applicable to an industrial/commercial-type siliconization process to improve water shedding properties. Finally, we have shown that our coating can be used as a first coating on a range of industrially-relevant materials to allow subsequent deposition of liquid-like coatings.

ASSOCIATED CONTENT

**Notes**

The authors declare no competing financial interest.

ACKNOWLEDGEMENTS

The authors thank Dr. Hongyu Zhao for assistance with ellipsometry measurements. We acknowledge the support of the UK Engineering and Physical Sciences Research Council through grants EP/V049348/1, EP/V049615/1 and EP/V049615/2.

REFERENCES

1. Funke, S. Cartridge Filling with Biopharmaceuticals with Focus on the Optimization of the Siliconization Process. (PhD Thesis, Ludwig-Maximilians-Universität München, Munich, 2016).

2. Solvias AG. Silicones for Pharmaceutical Applications – Analytical Capabilities at Solvias. *White Paper* Preprint at https://www.pharmaceutical-technology.com/downloads/whitepapers/clinical-trials/silicones-in-pharmaceutical-applications/ (2017).

3. SGD Pharma. Siliconization. A Unique Treatment Process Protecting and Adding Value to your Pharmaceutical Product. Preprint at https://www.sgd-pharma.com/siliconization-unique-treatment-process-protecting-and-adding-value-your-pharmaceutical-product (2021).

4. ELKEM. Improving Parenteral Packaging Functionality with Siliconization Silicones Healthcare. *White Paper* Preprint at https://magazine.elkem.com/healthcare/optimal-silicone-layer-on-parenteral-packaging/ (2022).

5. Colas, A., Siang, J. & Ulman, K. Silicones in Pharmaceutical Applications. Part 5: Siliconization of Parenteral Packaging Components. in 1–5 (Midland (MI): Dow Corning Corporation, 2006).

6. Petersen, C. & Zeiss, B. Syringe Siliconization. *International Pharmaceutical Industry* **7**, 78–84 (2015).

7. Reuter, B. & Petersen, C. Syringe Siliconization Trends. methods, analysis procedures. *TechnoPharm* **2**, 238–244 (2012).

8.  Melo, G. B. *et al.* Critical analysis of techniques and materials used in devices, syringes, and needles used for intravitreal injections. *Prog Retin Eye Res* **80**, 100862 (2021).

9.  Nelson, R. E., Hyun, D., Jezek, A. & Samore, M. H. Clinical Infectious Diseases Mortality, Length of Stay, and Healthcare Costs Associated With Multidrug-Resistant Bacterial Infections Among Elderly Hospitalized Patients in the United States. *Clinical Infectious Diseases* ® **74**, 1070–80 (2022).

10. Ikuta, K. S. *et al.* Global mortality associated with 33 bacterial pathogens in 2019: a systematic analysis for the Global Burden of Disease Study 2019. *The Lancet* **400**, 2221–2248 (2022).

11. Metcalf, D. & Bowler, P. Biofilm delays wound healing: A review of the evidence. *Burns Trauma* **1**, 5 (2013).

12. Butt, H. *et al.* Contact angle hysteresis. *Curr Opin Colloid Interface Sci* **59**, 101574 (2022).

13. McHale, G., Gao, N., Wells, G. G., Barrio-Zhang, H. & Ledesma-Aguilar, R. Friction Coefficients for Droplets on Solids: The Liquid–Solid Amontons' Laws. *Langmuir* **38**, 4425–4433 (2022).

14. Barrio-Zhang, H. *et al.* Contact-angle hysteresis and contact-line friction on slippery liquid-like surfaces. *Langmuir* **36**, 15094–15101 (2020).

15. Khatir, B., Azimi Dijvejin, Z., Serles, P., Filleter, T. & Golovin, K. Molecularly capped omniphobic polydimethylsiloxane brushes with ultra-fast contact line dynamics. *Small* **2301142**, 1–12 (2023).

16. Sharma, B. Immunogenicity of therapeutic proteins. Part 2: Impact of container closures. *Biotechnol Adv* **25**, 318–324 (2007).

17. Teska, B. M. Interactions between Therapeutic Protein Formulations and Surfaces. (PhD Thesis, University of Colorado, Denver, 2015).

18. Krayukhina, E., Tsumoto, K., Uchiyama, S. & Fukui, K. Effects of Syringe Material and Silicone Oil Lubrication on the Stability of Pharmaceutical Proteins. *J Pharm Sci* **104**, 527–535 (2015).

19. Fathi, F., Altiraihi, T., Mowla, S. J. & Movahedin, M. Formation of embryoid bodies from mouse embryonic stem cells cultured on silicon-coated surfaces. *Cytotechnology* **59**, 11–16 (2009).

20. Mundry, T., Surmann, P. & Schurreit, T. Surface characterization of polydimethylsiloxane treated pharmaceutical glass containers by X-ray-excited photo- and Auger electron spectroscopy. *Fresenius J Anal Chem* **368**, 820–831 (2000).

21. Rödel, E., Blatter, F., Büttücer, J.-P., Weirich, W. & Mahler, H.-C. Contact Angle Measurement on Glass Surfaces of Injection Solution Containers. *Pharmazeutische Industrie* **75**, 328–330 (2013).

22. Funke, S. *et al.* Optimization of the bake-on siliconization of cartridges. Part II: Investigations into burn-in time and temperature. *European Journal of Pharmaceutics and Biopharmaceutics* **105**, 209–222 (2016).

23. Onda, T., Shibuichi, S., Satoh, N. & Tsujii, K. Super-water-repellent fractal surfaces. *Langmuir* **12**, 2125–2127 (1996).

24. Neinhuis, C. & Barthlott, W. Characterization and distribution of water-repellent, self-cleaning plant surfaces. *Ann Bot* **79**, 667–677 (1997).

25. Barthlott, W. & Neinhuis, C. Purity of the sacred lotus, or escape from contamination in biological surfaces. *Planta* **202**, 1–8 (1997).

26. Quéré, D. Wetting and Roughness. *Annu Rev Mater Res* **38**, 71–99 (2008).

27. Wong, T.-S. *et al.* Bioinspired self-repairing slippery surfaces with pressure-stable omniphobicity. *Nature* **477**, 443–447 (2011).

28. Lafuma, A. & Quéré, D. Slippery pre-suffused surfaces. *Europhys Lett* **96**, 56001 (2011).

29. Krumpfer, J. W. & McCarthy, T. J. Rediscovering Silicones: "Unreactive" Silicones React with Inorganic Surfaces. *Langmuir* **27**, 11514–11519 (2011).

30. Wang, L. & McCarthy, T. J. Covalently attached liquids: Instant omniphobic surfaces with unprecedented repellency. *Angewandte Chemie International Edition* **55**, 244–248 (2016).

31. Chen, L., Huang, S., Ras, R. H. A. & Tian, X. Omniphobic liquid-like surfaces. *Nat Rev Chem* **7**, 123–137 (2023).

32. Gresham, I. J. & Neto, C. Advances and challenges in slippery covalently-attached liquid surfaces. *Adv Colloid Interface Sci* **315**, 102906 (2023).

33. McHale, G., Afify, N., Armstrong, S., Wells, G. G. & Ledesma-Aguilar, R. The Liquid Young's Law on SLIPS: Liquid–Liquid Interfacial Tensions and Zisman Plots. *Langmuir* **38**, 10032–10042 (2022).

34. Abbas, A., Wells, G. G., McHale, G., Sefiane, K. & Orejon, D. Silicone Oil-Grafted Low-Hysteresis Water-Repellent Surfaces. *ACS Appl Mater Interfaces* **15**, 11281–11295 (2023).

35. Zhang, L., Guo, Z., Sarma, J. & Dai, X. Passive Removal of Highly Wetting Liquids and Ice on Quasi-Liquid Surfaces. *ACS Appl Mater Interfaces* **12**, 20084–20095 (2020).

36. Khatir, B., Shabanian, S. & Golovin, K. Design and High-Resolution Characterization of Silicon Wafer-like Omniphobic Liquid Layers Applicable to Any Substrate. *ACS Appl Mater Interfaces* **12**, 31933–31939 (2020).

37. Zhao, X., Khandoker, M. A. R. & Golovin, K. Non-Fluorinated Omniphobic Paper with Ultralow Contact Angle Hysteresis. *ACS Appl Mater Interfaces* **12**, 15748–15756 (2020).

38. Halvey, A. K. *et al.* Rapid and Robust Surface Treatment for Simultaneous Solid and Liquid Repellency. *ACS Appl Mater Interfaces* **13**, 53171–53180 (2021).

39. Armstrong, S. *et al.* Pinning-Free Evaporation of Sessile Droplets of Water from Solid Surfaces. *Langmuir* **35**, 2989–2996 (2019).

40. Young, T. An Essay on the Cohesion of Fluids. *Proceedings of the Royal Society of London* **1**, 171–172 (1800).

41. Zhu, Y. *et al.* Slippery Liquid-Like Solid Surfaces with Promising Antibiofilm Performance under Both Static and Flow Conditions. *ACS Appl Mater Interfaces* **14**, 6307–6319 (2022).

42. Gresham, I. J., Lilley, S. G., Nelson, A. R. J., Koynov, K. & Neto, C. Nanostructure Explains the Behavior of Slippery Covalently Attached Liquid Surfaces. *Angewandte Chemie International Edition* **62**, 1–15 (2023).

43. Zhang, L., Guo, Z., Sarma, J. & Dai, X. Passive Removal of Highly Wetting Liquids and Ice on Quasi-Liquid Surfaces. *ACS Appl Mater Interfaces* **12**, 20084–20095 (2020).

44. Vahabi, H. *et al.* Designing non-textured, all-solid, slippery hydrophilic surfaces. *Matter* **5**, 4502–4512 (2022).

45. Papra, A., Gadegaard, N. & Larsen, N. B. Characterization of Ultrathin Poly(ethylene glycol) Monolayers on Silicon Substrates. *Langmuir* **17**, 1457–1460 (2001).

46. Kaneko, S., Urata, C., Sato, T., Hönes, R. & Hozumi, A. Smooth and Transparent Films Showing Paradoxical Surface Properties: The Lower the Static Contact Angle, the Better the Water Sliding Performance. *Langmuir* **35**, 6822–6829 (2019).

47. de Gennes, P.-G., Brochard-Wyart, F. & Quéré, D. *Capillarity and Wetting Phenomena. Capillarity and Wetting Phenemona: Drops, Bubbles, Pearls, and Waves* (Springer, New York, 2004). doi:10.1007/978-0-387-21656-0.

48. Zhao, H. *et al.* Scalable Slippery Omniphobic Covalently Attached Liquid Coatings for Flow Fouling Reduction. *ACS Appl Mater Interfaces* **13**, 38666–38679 (2021).

49. Jiao, S. *et al.* A simple and universal strategy for liquid-like coating suitable for a broad range of liquids on diverse substrates. *Cell Rep Phys Sci* **4**, (2023).

50. McHale, G. *et al.* Transforming Auxetic Metamaterials into Superhydrophobic Surfaces. *Small Struct* **5**, 2300458 (2024).

# Supplementary information: Transforming Siliconization into Slippery Liquid-like Coatings

*Hernán Barrio-Zhang[1$], Glen McHale[1*], Gary G. Wells[1], Rodrigo Ledesma-Aguilar[1], Rui Han[2], Nicholas Jakubovics[3], Jinju Chen[2].*

[1]Institute for Multiscale Thermofluids, School of Engineering, University of Edinburgh, The King's Buildings, Mayfield Road, Edinburgh EH9 3FB, UK. [2]Department of Materials, Loughborough University, Loughborough, LE11 3TU, UK. [3]School of Dental Sciences, Faculty of Medical Sciences, Newcastle University, Newcastle Upon Tyne, NE2 4BW, UK.

*Email: glen.mchale@ed.ac.uk

$Email: hbarrio@exseed.ed.ac.uk

**Comparison of a droplet sliding on SOCAL and capped SOCAL on a glass substrate.**

Figure S1 shows video still images of a 20 µL DI water droplet sliding at a tilt angle of 30° on SOCAL (top panels) and on capped SOCAL (bottom panels). The difference in sliding velocity is dramatic, showing up to a ~100-fold increase in velocity for the capped sample in some cases.

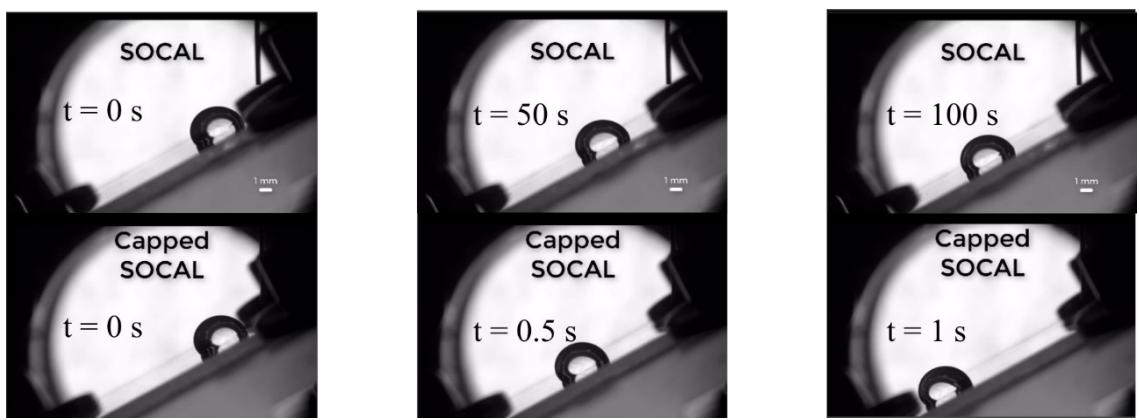

Figure S1. SOCAL vs Capped SOCAL droplet sliding. The still images in time (t) show that the sliding velocity on SOCAL surfaces is slow, where the droplet moves partially along the surface over the course of 100 s. In contrast, on the capped SOCAL, the sliding velocity of the droplet is dramatically increased, traversing along the surface in 1 s.

**Damage of SoSLIC coating on Stainless steel sample post QLS deposition**

After attempting to deposit liquid-like coatings using vapour deposition (QLS procedure) on Stainless Steel SoSLIC, the results did not bear a positive outcome since the process damaged the intermediate layer, as can be seen in Figure S2. The SoSLIC layer is completely destroyed and peel-off is observed. The damage is likely due to high concentrations of HCl produced in the QLS reaction.

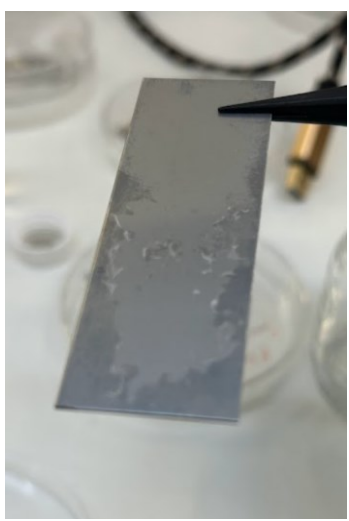

Figure S2. Coating destroyed post QLS treatment. This image shows the damage caused by the QLS treatment on an SoSLIC coated stainless steel sample.